\documentclass[prd,nofootinbib,preprint]{revtex4}

\usepackage{subfigure}
\usepackage{graphicx}
\usepackage{pstricks}
\usepackage{latexsym}

\begin{document}

\title{Impact of right-handed interactions on the propagation
of Dirac and Majorana neutrinos in matter}
\vspace*{1.5cm}

\author{F. del Aguila,}
\email{faguila@ugr.es}
\affiliation{%
Departamento de F{\'\i}sica Te\'orica y del Cosmos and \\
Centro Andaluz de F{\'\i}sica de Part{\'\i}culas Elementales (CAFPE), \\
Universidad de Granada, E-18071 Granada, Spain}

\author{J. Syska}
\email{jacek@server.phys.us.edu.pl}
\author{M. Zra{\l}ek}
\email{zralek@us.edu.pl}
\affiliation{%
Department of Field Theory and Particle Physics, Institute of Physics, \\
University of Silesia, Uniwersytecka 4, 40-007 Katowice, Poland}
\vspace*{1cm}

\begin{abstract}
\vspace*{0.5cm}
Dirac and Majorana neutrinos can be distinguished in relativistic
neutrino oscillations if new right-handed interactions exist,
due to their different propagation in matter.
We review how these new interactions affect neutrino oscillation
experiments and discuss the size of this eventually observable
effect for different oscillation channels, baselines and neutrino
energies.
\end{abstract}

\preprint{}


\maketitle

\section{Introduction}
\label{intr}

One of the fundamental mysteries of neutrino
physics is the nature of massive neutrinos \cite{mass}.
They can have Dirac or Majorana masses \cite{Majorana},
with their own signatures in each case.
In particular, it is assumed that
the best way to establish their character is to search for
neutrinoless double $\beta$ decay $((\beta\beta)_{0\nu})$
\cite{Bilenky_Petkow_Kayser, Mohapatra-Pal-book,
Bilenky1_Barabash_Valle_Petcov}. Indeed, its mere
observation would imply that neutrinos are Majorana particles.
However, if neutrino masses are a fraction of eV
and no New Physics (NP) gives large enough contributions,
it will be quite difficult to observe $(\beta\beta)_{0\nu}$
\cite{Lyman_Haxby, Abela, Maltoni, DKGZ}.

Although oscillation experiments are not
sensitive to the absolute scale of neutrino masses, they are able
to probe very tiny neutrino mass differences
\cite{Gonzales_Maltoni, sno, Fogli}. Hence, the obvious question
is if neutrino oscillations can distinguish between Dirac and
Majorana neutrino masses \cite{Pontecorvo, Czakon_Gluza_Zralek}.
This is not possible in neutrino oscillations in vacuum
\cite{Bilenky3, Doi}. The reason is that the additional
phases in the neutrino mixing matrix distinguishing Majorana from
Dirac neutrinos disappear from the oscillation probability
expressions.
The oscillation in matter can differ from that in vacuum,
leading to both the change of the effective mixing angles
and of the effective masses
\cite{Wolfenstein_Mikheev_Smirnov, Rosen_Kuo, Aguila_Zralek}.
However, if neutrinos are relativistic and only interact
through left-handed (LH) currents, the oscillation in matter
does not discriminate between both types of fermions either.
This is not the case for non-relativistic neutrinos
\cite{Kiers_Weiss} but unfortunately these neutrinos are not
available in any experiment, at least up to now.
So, within the New Standard Model ($\nu SM $) \cite{Boehm_Vogel}
and for neutrinos in the mass range of tens of meV it will be very
difficult to find experimental evidence which can distinguish
Dirac from Majorana neutrinos.

Then, one may wonder if there is some type of NP which can modify
neutrino propagation in a dense medium
\cite{NSI,ggetal,fmbv,garbutt,Winter}
to allow for the determination
of the neutrino mass character.
As we will review, although NP can give rise to a
low energy Hamiltonian with any tensor structure for neutrino
bilinears $\bar{\nu} \Gamma ^a \nu$, with
$\Gamma^{a} = I,\  \gamma_{5},\  \gamma^{\mu}, \
\gamma^{\mu} \gamma_{5}$, and $\sigma^{\mu\nu}$, the scalar and
pseudoscalar currents vanish or are suppressed for relativistic
neutrinos. They are $0(\frac{m}{E})$, where $m$ is a light
neutrino mass and $E$ the neutrino energy. Vector and axial-vector helicity
flip and tensor helicity non-flip transitions have similar
suppression factors. Only the vector and axial-vector helicity
non-flip transitions and the tensor helicity flip ones are
unsuppressed, and may then help to discriminate between relativistic
Dirac and Majorana neutrinos propagating in a dense medium. Here we will
concentrate on the spin non-flip vector and axial-vector transitions.
Tensor interactions have been discussed in this
context in \cite{Bergmann 1999}. In any case, as the production
and detection of neutrinos (antineutrinos) with positive
(negative) helicity are strongly suppressed, the sensitivity to
spin-flip tensor transitions is reduced. Besides, they require a
polarised background to manifest, and if generated at
higher orders in perturbation theory, they are also
extra suppressed by loop factors \footnote{They also result
from Fierz rearrangement of (pseudo)scalar interactions and
can be of the same order as the other effective
four-fermion interactions, being only suppressed by
the corresponding inverse power of the large effective mass scale
and eventually by the product of small (Yukawa) couplings.}
\cite{effectivelagrangian}.

In this paper we will consider an effective Lagrangian with
only arbitrary vector and axial-vector couplings modifying the LH
structure of the $\nu SM $.
This is enough to generate (new) vector and axial-vector four-fermion
interactions which are those to be probed by relativistic neutrinos.
The new couplings affect differently
the propagation of Dirac and Majorana neutrinos in matter.
The largest effect results from the addition of new right-handed (RH)
neutral currents,
being the corresponding transition probability dependence linear
\footnote{Although leading order terms in the effective Lagrangian
expansion may be actually subdominant in definite models.}.
We discuss its size for different channels as a function
of the neutrino energy and the baseline, showing that it can
eventually allow for discriminating between Dirac and Majorana
neutrinos.

We first revise the necessary known results on neutrino
propagation in a dense medium.
In next Section we describe the neutrino interaction with matter
introducing a general, relativistic invariant Hamiltonian, and
show that for relativistic neutrinos only vector, axial-vector and
tensor terms have contributions to neutrino oscillation amplitudes
which are not suppressed by $\frac{m}{E}$ factors. In Section
\ref{Lagrangians and Hamiltonian densities} we
discuss the Lagrangian with general LH and RH currents. As
deviations from the $\nu SM $ are strongly constrained by existing
experimental data, they must be small. Then, using the
corresponding four-fermion effective Hamiltonian we calculate the
effective interaction with matter for Dirac and Majorana
neutrinos. In Section \ref{Propagation} the probabilities for neutrino
oscillations in matter are derived for a neutral background with uniform
spin and momentum distributions \cite{Gluza_Zralek}. We
present our numerical results for neutrino propagation inside
Earth for different channels, baselines and neutrino energies
in Section \ref{bounds}.
Finally, last Section is devoted to conclusions and to comment on
some necessary ingredients of more fundamental models with RH
neutrino interactions.

\section{Dirac versus Majorana neutrinos}
\label{Dirac or Majorana}

Let us evaluate the suppression factors for
the different transitions.
The effective Hamiltonian describing the
coherent neutrino scattering in a dense medium has the form
\begin{eqnarray}
\label{H-eff-nu-copy}
{\cal H}^{int} =
\sum_{a} \sum_{i,j} \left[ (z_{a})_{ij} \;
(\bar{\nu}_{i} \Gamma^{a} \nu_{j}) + (z_{a}^{*})_{ij} \;
(\bar{\nu}_{j} \overline{\Gamma}^{a} \nu_{i}) \right] \, ,
\end{eqnarray}
where $\Gamma^{a} = I,\  \gamma_{5},\  \gamma^{\mu}, \
\gamma^{\mu} \gamma_{5}$ and $\sigma^{\mu\nu} =
\frac{i}{2}[\gamma^{\mu},\gamma^{\nu}]$, and $z_{a}$ are the
complex scalar ($z_{S}$), pseudoscalar ($z_{P}$), vector
$(z_{V\mu})$, axial-vector ($z_{A\mu}$) and tensor $(z_{T
\mu\nu})$ matrices characterizing the background.
The sum $i,j$ runs over mass eigenstates.
For the moment it is enough to know that the Hamiltonian
is hermitian, what follows from the relations
\begin{eqnarray}
\label{Lambda}
\overline{\Gamma}^{a} \equiv \gamma_{0}\
(\Gamma^{a}) ^{\dag}\
\gamma_{0} = \left\{
\begin{array}{lll}  &\Gamma^{a}& \; ,
\;\;\; {\rm for} \;\; a= S, \; V, \; A ,\; T , \vspace{1mm} \\
 - & \Gamma^{a}&   \; , \;\;\; {\rm for} \;\; a = P \; .
\end{array} \right.
\end{eqnarray}
The $j \rightarrow i$ matrix element
\begin{eqnarray*}
\label{n-Lambda-n}
{\cal H}_{ij} = \langle
\nu_{i} | {\cal H}^{int} | \nu_{j} \rangle
\end{eqnarray*}
can be calculated using the plane wave decomposition for the
neutrino fields and contracting the corresponding creation and
annihilation operators with those of the initial and final
Dirac neutrino, Dirac antineutrino or Majorana neutrino states.
We obtain for each case
\begin{eqnarray}
\label{amplitute n m}
\begin{array}{lll}
{\cal H}^{D}_{ij} & = & \sum_{a} [ (z_{a})_{ij}\ {\bar
u}_{i}\Gamma^{a} u_{j} +(z_{a}^{*})_{ji}\ {\bar u}_{i}
{\overline{\Gamma}}^{a} u_{j} ] \; , \vspace{3mm} \\
{\cal H}^{D}_{{\overline i}{\overline j}} & = & - \sum_{a}
[ (z_{a})_{ji}\ {\bar v}_{j}\Gamma^{a} v_{i} +
(z_{a}^{*})_{ij} \, {\bar v}_{j} \bar{\Gamma}^{a} v_{i} ] \; ,
\vspace{3mm} \\
{\cal H}^{M}_{ij} & = & {\cal H}^{D}_{ij} +
{\cal H}^{D}_{{\overline i}{\overline j}} \; ,
\end{array}
\end{eqnarray}
respectively.
Yet it is more convenient to rewrite them only as a function
of positive frequency spinors using the charge conjugation
relation $v_{i} = C \, \bar{u}_{i}^{T}$, together with
Eq. (\ref{Lambda}) and the relations
\begin{eqnarray}
\label{Lambda C}
C \ ({\Gamma}^{a}) ^{T} \ C^{-1} = \left\{
\begin{array}{lll}  &\Gamma^{a}&,
\;\;\; {\rm for} \;\; a = S,\ P,\ A , \vspace{1mm} \\
 - & \Gamma^{a}&, \;\;\; {\rm for} \;\; a =  V, \; T .
\end{array} \right.
\end{eqnarray}\\
Then, they read
\begin{eqnarray}
\label{M-i-k-Dirac}
\begin{array}{lll}
{\cal H}^{D}_{ij} & = & (Z_{S})_{ij} \,
{\bar u}_{i} u_{j} + (Z_{P})_{ij} \, {\bar u}_{i} \gamma_{5} u_{j}
+ (Z_{V \mu})_{ij} \, {\bar u}_{i} \gamma^{\mu} u_{j} \vspace{1.5mm} \\
& & + (Z_{A \mu})_{ij} \, {\bar u}_{i} \gamma^{\mu}
\gamma_{5} u_{j} + (Z_{T \mu\nu})_{ij} \  {\bar u}_{i}
\sigma^{\mu\nu} u_{j} \; , \vspace{3mm} \\
{\cal H}^{D}_{{\overline i}{\overline j}}
& = & (Z_{S}^{*})_{ij} \, {\bar u}_{i} u_{j} - (Z_{P}^{*})_{ij} \,
{\bar u}_{i} \gamma_{5} u_{j} - (Z_{V \mu}^{*})_{ij} \, {\bar u}_{i}
\gamma^{\mu} u_{j} \vspace{1.5mm} \\
& & + (Z_{A \mu}^{*})_{ij} \, {\bar u}_{i} \gamma^{\mu} \gamma_{5}
u_{j} - (Z_{T \mu\nu}^{*})_{ij} \  {\bar u}_{i} \sigma^{\mu\nu} u_{j} \; ,
\vspace{3mm} \\
{\cal H}^{M}_{ij} & = & 2 \, (Re Z_{S})_{ij} \,
{\bar u}_{i} u_{j} + 2 \, i \, (Im Z_{P})_{ij} \, {\bar u}_{i} \gamma_{5} u_{j}
+ 2 \, i \, (Im Z_{V \mu})_{ij} \, {\bar u}_{i} \gamma^{\mu} u_{j} \vspace{1.5mm} \\
& & + 2 \, (Re Z_{A \mu})_{ij} \, {\bar u}_{i} \gamma^{\mu} \gamma_{5} u_{j}
+ 2 \, i \, (Im Z_{T \mu\nu})_{ij} \  {\bar u}_{i} \sigma^{\mu\nu} u_{j} \; ,
 \\
\end{array}
\end{eqnarray}
where the capital $Z$ matrices are constructed from the small $z$ ones.
They are hermitian $Z = z\ +\  z^{\dagger}$ for S,V,A,T and
antihermitian $Z = z\ -\  z^{\dagger}$ for P terms.
As we shall see, Eq. (\ref{M-i-k-Dirac}) is further simplified
for relativistic neutrinos.

\subsection{Relativistic limit}
\label{Rel-limit}

In the Weyl representation the spinor $u_{\lambda}(p_{\nu})$ for a
particle with helicity ${\frac{\lambda}{2}}$, energy $E_{\nu}$ and
momentum $p_{\nu}$, and normalized in such a way that $u^{\dag} u
= 1$, has the form
\begin{eqnarray}
\label{u and v}
u_{\lambda}(p_{\nu}) = \left(
\matrix{ \sqrt{\frac{E_{\nu} + \lambda |{\vec p}_{\nu}| }
{2 E_{\nu}}} {\chi}_{\lambda} \vspace{3mm} \cr \sqrt{\frac{E_{\nu} -
\lambda |{\vec p}_{\nu}| }{2 E_{\nu}}} {\chi}_{\lambda} \cr}
\right) \; ,
\end{eqnarray}
where $\chi_{\lambda}$ is the Pauli helicity spinor properly
normalized $\chi^{\dag}\chi = 1$. In order to compare the spinor
products for relativistic neutrinos with different masses, it is
convenient to use a common energy, which is independent of
the particular small neutrino masses involved. Indeed, noting that
in forward neutrino scattering all momenta are parallel
$\overrightarrow{p}_{i} = p_{i} \overrightarrow{i}$ and following
\cite{Giunti Kim}, we can write the energy and the momentum of any
light neutrino
\begin{eqnarray}
\label{u and v}
E_{i} = E + \xi \frac{m_{i}^{2}}{2E} \; ,
\ \ \ \ \ p_{i} = E - (1-\xi)
\frac{m_{i}^{2}}{2E} \; ,
\end{eqnarray}
where $\xi$ is some parameter which depends on the production or
detection process, and $E$ is the neutrino energy for zero
neutrino mass. Using these expressions we can evaluate
the spinor products entering in Eq. (\ref{M-i-k-Dirac})
\begin{eqnarray}
\label{uu-rel}
\begin{array}{lcl}
{\bar u}_{i}(\lambda) u_{j}(\lambda)  &=&
\frac{m_{i} + m_{j}}{2 E} + 0((\frac{m}{E})^{3}) ,
\vspace{2mm} \\
{\bar u}_{i}(\lambda) u_{j}(- \lambda)  &=& 0 ,
\vspace{4mm} \\
{\bar u}_{i}(\lambda) \gamma^{5} u_{j}(\lambda) &=&
\lambda\ \frac{m_{i} - m_{j}}{2 E} + 0((\frac{m}{E})^{3}) ,
\vspace{2mm} \\
{\bar u}_{i}(\lambda) \gamma^{5} u_{j}(-\lambda) &=& 0 ,
\vspace{4mm} \\
{\bar u}_{i}(\lambda) \gamma^{\mu} u_{j}(\lambda) &=& n^{\mu} +
0((\frac{m}{E})^{2}) ,
\vspace{2mm} \\
{\bar u}_{i}(-1) \gamma^{\mu} u_{j}(+1) &=& - [{\bar u}_{i}(+1)
\gamma^{\mu} u_{j}(-1)]^{*} = m^{\mu} \ \frac{m_{i} - m_{j}}{2 E}
+ 0((\frac{m}{E})^{3}) ,
\vspace{4mm} \\
{\bar u}_{i}(\lambda) \gamma^{\mu} \gamma_{5} u_{j}(\lambda) &=&
\lambda\ n^{\mu}\ + 0((\frac{m}{E})^{2}) ,
\vspace{2mm} \\
{\bar u}_{i}(-1) \gamma^{\mu} \gamma_{5} u_{j}(+1) &=&
[{\bar u}_{i}(+1) \gamma^{\mu} \gamma_{5}
 u_{j}(-1)]^{*} = m^{\mu} \ \frac{m_{i} + m_{j}}
{2 E} + 0((\frac{m}{E})^{3}) ,
\vspace{4mm} \\
{\bar u}_{i}(\lambda) \sigma^{0k} u_{j}(\lambda) &=& i \ n^{k}\
\frac{m_{i} - m_{j}}{2
 E} + 0((\frac{m}{E})^{3}) ,
\vspace{2mm} \\
{\bar u}_{i}(\lambda) \sigma^{kl} u_{j}(\lambda) &=& \lambda \
\varepsilon^{klr} n^{r}\ \frac{m_{i} + m_{j}}{2
 E} + 0((\frac{m}{E})^{3}) ,
\vspace{2mm} \\
{\bar u}_{i}(-1) \sigma^{0k} u_{j}(+1) &=&
[{\bar u}_{i}(+1) \sigma^{0k} u_{j}(-1)]^{*} =
 i \, m^{k} + 0((\frac{m}{E})^{2}) ,
\vspace{2mm} \\
{\bar u}_{i}(-1) \sigma^{kl} u_{j}(+1) &=& [{\bar u}_{i}(+1)
\sigma^{kl} u_{j}(-1)]^{*} =
 \varepsilon^{klr} m^{r} + 0((\frac{m}{E})^{2}) ,
\\
\end{array}
\end{eqnarray}
where $n^{\mu} = (1,\overrightarrow{n})$ with $\overrightarrow{n}=
(\sin\theta \cos\varphi, \sin\theta \sin\varphi,\cos\theta)$
the direction of the neutrino momentum, and $m^{\mu} =
(0,\overrightarrow{m})$ with $\overrightarrow{m}= ( \cos\theta
\cos\varphi - i \sin\varphi, \cos\theta \sin\varphi+ i
\cos\varphi, - \sin\theta)$ orthogonal to
$\overrightarrow{n}$ and defined up to an unphysical global phase;
whereas $\varepsilon^{klr}$ is the totally antisymmetric tensor
with $\varepsilon^{123} = 1$.

Hence, we see that in the relativistic limit for forward neutrino
scattering the scalar and pseudoscalar terms can be effectively
omitted.  If in addition we assume that the tensor term is negligible
(the general case including non-vanishing tensor interactions will be
considered elsewhere), then not only the corresponding suppressed
spin non-flip but the spin flip terms can be omitted,
and Eq. (\ref{M-i-k-Dirac}) reduces to
\begin{eqnarray}
\label{M-i-k-Dirac-rel 1}
\begin{array}{lll}
{\cal H}^{D}_{ij}(\lambda)
&=& \left( Z_{V \mu} + \lambda Z_{A \mu} \right)_{ij}\
 n^{\mu} \; , \vspace{2mm} \\
{\cal H}^{D}_{{\overline i}{\overline j}}(\lambda) &=&  \left(
- Z_{V \mu}^{*} + \lambda Z_{A \mu}^{*} \right)_{ij} \, n^{\mu} \; ,
\vspace{2mm} \\
{\cal H}^{M}_{ij}(\lambda)
&=&  2 \left( i \, Im Z_{V \mu} +
\lambda \, Re Z_{A \mu} \right)_{ij}\, n^{\mu}  \; .
\end{array}
\end{eqnarray}
Thus, the allowed transitions do not flip helicity and depend
on the vector and axial-vector couplings only.
In order to answer the question on the nature of the neutrino mass
we must compare the amplitudes
${\cal H}^{D}_{ij}(\lambda = -1)$ with
${\cal H}^{M}_{ij}(\lambda = -1)$ for particles, and
${\cal H}^{D}_{{\overline i}{\overline j}}(\lambda = + 1)$
with ${\cal H}^{M}_{ij}(\lambda = + 1)$
for antiparticles.

In the $\nu SM $, where neutrinos interact only through LH currents,
the vector and axial-vector $Z$ matrices are related,
$Z_{V\mu} = - Z_{A\mu}$, and the only non-vanishing transitions read
\begin{eqnarray}
\label{M-i-k-Dirac-rel-L-R}
\begin{array}{l}
{\cal H}^{D}_{ij}(\lambda
= - 1) = {\cal H}^{M}_{ij}(\lambda = - 1) = 2 (Z_{V
\mu})_{ij}\, n^{\mu},\; \vspace{3mm} \\
{\cal H}^{D}_{{\overline i}{\overline j}}(\lambda = +1) =
{\cal H}^{M}_{ij}(\lambda = +1) =
- 2 (Z_{V\mu}^{*})_{ij} \, n^{\mu}. \;
\end{array}
\end{eqnarray}
In any model where neutrinos besides interact through RH currents
$Z_{V \mu} \neq - Z_{A \mu}$, and there are deviations
from the $\nu SM $ predictions in Eq. (\ref{M-i-k-Dirac-rel-L-R})
which also affect differently the Dirac and Majorana neutrino
propagation.
Indeed, Eq. (\ref{M-i-k-Dirac-rel 1}) gives our main input
\begin{eqnarray}
\label{D-M-rel-(-1)}
\begin{array}{l}
{\cal H}^{D}_{ij}(\lambda = - 1)
- {\cal H}^{M}_{ij}(\lambda = - 1) = \left( Z_{V
\mu}^{*} +  Z_{A\mu}^{*}\right)_{ij} \, n^{\mu} \; , \vspace{3mm} \\
{\cal H}^{D}_{{\overline i}{\overline j}}
(\lambda = +1) - {\cal H}^{M}_{ij}(\lambda = +1)
=  -  \left( Z_{V\mu} +  Z_{A\mu}\right)_{ij} \, n^{\mu} \; .
\end{array}
\end{eqnarray}
Hence, even in the relativistic limit there is in principle the possibility
of distinguishing between the Dirac and Majorana neutrinos, but only if
besides the dominant LH current there is some RH piece.
Whether this can have practical consequences, it depends on the
strength of the RH interaction.

\section{Lagrangian and effective low-energy Hamiltonian densities}
\label{Lagrangians and Hamiltonian densities}

One may adopt a more phenomenological approach at this point
and parameterise the difference between Dirac and Majorana amplitudes
in Eq. (\ref{D-M-rel-(-1)}), calculating the corresponding oscillation
probabilities afterwards and looking for the largest possible
effects. However, one must also worry about the physical
implications of such a parameterisation and the experimental
constraints that restrict the different parameters.
Then, it is more enlightening to start extending the
$\nu SM$ Lagrangian describing the coherent scattering of
neutrinos on background fermions $f$,
\begin{eqnarray*}
\label{n-f} \nu_j + f \rightarrow \nu_i + f \; .
\end{eqnarray*}
Before discussing any specific model, let us introduce an effective
Lagrangian with arbitrary LH and RH currents and show that the largest
effects are associated to new RH neutral currents involving the light
neutrinos.

\subsection{The effective left-right interaction Lagrangian}

The $\nu SM$ charged current Lagrangian can be
generalised to include new LH and RH couplings
\begin{eqnarray}
\label{L-int-CC-general}
{\cal L}_{CC} = - \frac{e}{2 \sqrt{2}
\sin{\theta_{W}}} \sum_{i,\alpha} \bar{\nu}_{i} \gamma^{\mu} \left[
\varepsilon^{C}_{L} (1 - \gamma_{5}) + \delta^{C}_{R} (1 +
\gamma_{5}) \right] \, U^{*}_{\alpha i } \, l_{\alpha} \,
W_{\mu}^{+} + h.c. \; ,
\end{eqnarray}
where $\varepsilon^{C}_{L}$ and $\delta^{C}_{R}$ are taken to
be global factors deviating slightly from their $\nu SM$ values, 1 and
0, respectively. This is enough for as we will show, they enter
quadratically in the effective interaction Hamiltonian. So,
$|\delta^{C}_{R}|^2$ contributions are negligible because we
are interested in large (linear) effects.
Then, although in contrast with Eq. (\ref{L-int-CC-general})
the RH and LH charged currents can have in general different mixing
matrices, its explicit form will not matter in the RH case.
While $|\varepsilon^{C}_{L}|^2$ terms contribute the same to the
propagation of Dirac and Majorana neutrinos.
Thus, we will take $\varepsilon^{C}_{L} = 1$ and
$U_{\alpha i}$ to be the $3\times 3$ unitary matrix
giving the linear combinations of mass eigenstates with
well-defined lepton flavour \cite{MNS,Kim_Pevsner}
\begin{eqnarray*}
\label{n-ni}
\nu_{\alpha} = \sum_{i} U_{\alpha i} \, \nu_{i} \; .
\end{eqnarray*}
One must keep in mind, however, that in definite models the
terms neglected can be of the same order as those distinguishing
between Dirac and Majorana neutrinos below.

Similarly, the neutral current Lagrangian can be written
\begin{eqnarray}
\label{L-int-NC-general}
{\cal L}_{NC} &=& - \frac{e}{4
\sin{\theta_W} \cos{\theta_W}} \{ \sum_{i,j} \bar{\nu}_{i}
\gamma^{\mu} \left[ \varepsilon^{N \nu}_{L} (1 -
\gamma_{5})\delta_{ij} + \delta^{N \nu}_{R}
(1 + \gamma_{5})\Omega^R_{ij}\right]  {\nu}_{j}  + \nonumber \\
&+& \sum_{f} \bar{f} \gamma^{\mu} \left[
\varepsilon^{N f}_{L} (1 - \gamma_{5}) + \varepsilon^{N f}_{R} (1
+ \gamma_{5}) \right] f \} Z_{\mu} \; ,
\end{eqnarray}
where as before $\varepsilon^{N \nu}_{L}$ is a global factor of
order 1, its $\nu SM$ value, to keep track of the order of the
different terms. The LH piece for light neutrinos can be assumed
to be diagonal (and $\varepsilon^{N \nu}_{L} = 1$)
because it will not help to determine their
nature
\footnote{In models with neutrino mixing with
$\nu SM$ singlets there are (off-diagonal) terms
proportional to this mixing and coupling mass
eigenstates mainly participating of standard
neutrinos and of $\nu SM$ singlets, respectively;
as well as quadratic terms correcting the standard
couplings.}.
What is crucial is the general form of the RH
piece, that we parameterised with a global factor $\delta^{N
\nu}_{R}$ of order 0, the value in the $\nu SM$,
multiplying an otherwise arbitrary hermitian matrix $\Omega^R_{ij}$.
The background fermions can be taken to be $f
= e, u, d$ or $e, p ,n$, with their neutral couplings
\begin{eqnarray}
\label{c in SM}
\varepsilon_{L}^{N f} = 2 T^3_{f} -  q_{f} \sin^{2}\theta_{W} +
\delta_{L}^f \; ,
\;\;\;\; \varepsilon_{R}^{N f}
= - q_{f} \sin^{2}\theta_{W} + \delta_{R}^f \; ,
\end{eqnarray}
where $q_{f} \ (T^3_{f})$ is the fermion charge (third component
of the weak isospin) and $\delta_{L, R}^f$ are possible small
deviations from their standard values.

The new RH coupling in Eq. (\ref{L-int-NC-general})
requires the addition of RH neutrinos transforming
non-trivially under the $\nu SM$ gauge group, or the $Z$ mixing
with an extra gauge boson coupling to the new RH neutrinos if these are
$\nu SM$ singlets. Both scenarios have further consequences.
For example, RH neutrinos in non-trivial $SU(2)_L$ representations imply
new charged leptons which have not been observed, and the $Z$ mixing
with an extra gauge boson is strongly constrained by processes
not involving neutrinos. Definite models must evade these constraints.
This is not a problem in the Dirac case,
but the observed neutrinos
must mix with the new light neutrinos entering in the RH neutral
current if the Majorana neutrinos have to feel a different interaction
when travelling through matter.
Indeed, if there are only three light neutrinos feeling the $\nu SM$
interactions, their nature can not be
established in oscillation experiments with very energetic
neutrinos, although there are new sectors beyond the $\nu SM$.
In order to manifest the Majorana character the
light neutrinos must participate of new interactions and
then of new degrees of freedom
\footnote{Dirac neutrinos require RH counterparts, but it is
not required that these have other interactions.}.
In either case light neutrinos can mixed with new heavy neutrinos
\cite{Bekman 2002,collider}, but this mixing which must be
rather small is of no relevance to decide about the
nature of light neutrinos in oscillation experiments.
What matters is the mixing with new light neutrinos
with RH interactions. The experimental limits on such a
mixing are not so stringent, because all light neutrinos
are produced in the standard decay processes and there is no
deficit relative to the corresponding $\nu SM$ prediction.
On the other hand, in loop processes we have also to sum
over all light degrees of freedom, and the new contributions
are proportional to the (new) light neutrino mass differences.

We could also think in adding new (pseudo)scalar interactions
to generate an effective low energy Hamiltonian with RH neutrino
currents coupled to background fermions, but they are also strongly
constrained if, as we need, the new interaction couples the
observed neutrinos to matter fermions. Besides,
some (neutrino) mixing with new fermionic degrees of freedom
and between different spin 0 bosons are still necessary.

\subsection{The general effective interaction Hamiltonian}

The effective low energy four-fermion Hamiltonian
resulting from the former charged and neutral interaction Lagrangians
has the general form
\begin{eqnarray}
\label{H-eff-sum}
{ \cal{H}}_{eff} = \sum_{f=e,p,n}
\frac{G_{F}}{\sqrt{2}}  \sum_{i,j} \; \sum_{\, a=V, A}
\left( \bar{\nu}_{i} \Gamma^{a} \nu_{j} \right) \left[ \bar{f} \,
\Gamma_{a} \left( g_{f a}^{ij} + {\bar g}_{f a}^{ij} \gamma_{5}
\right)  f \, \right] \; ,
\end{eqnarray}
where we neglect the scalar and pseudoscalar terms generated by
Fierz rearrangement because for relativistic neutrinos they vanish
or are suppressed by small $\frac{m}{E}$ factors, as discussed in
Section \ref{Dirac or Majorana}.
Thus, we are only left with vector and axial-vector interactions,
$\Gamma^{a} = \gamma^{\mu}$, $\gamma^{\mu} \gamma_{5}$, with
couplings $g_{f a}^{ij}$ and ${\bar g}_{f a}^{ij}$ given
by
\begin{eqnarray}
\begin{array}{ll}
\label{g-c relations1}
g_{f V}^{ij} = g_{f L}^{ij} + g_{f R}^{ij}
\; , \;\;\;\;\; &  {\bar g}_{f V}^{ij} = {\bar g}_{f L}^{ij} + {\bar
g}_{f R}^{ij} \; ,  \vspace{3mm} \\
g_{f A}^{ij} =  - {\bar g}_{f
L}^{ij} + {\bar g}_{f R}^{ij} \; , \;\;\;\;\; & {\bar g}_{f A}^{ij}
= - g_{f L}^{ij} + g_{f R}^{ij} \; ,
\end{array}
\end{eqnarray}
where we introduce LH and RH couplings for later convenience, and
\begin{eqnarray}
\begin{array}{ll}
\label{g-c relations2}
g_{f L}^{ij} = \; ( A^{LL} + A^{LR} )^{f}_{ij}
\; , \;\;\;\;\; & {\bar g}_{f L}^{ij} = (- A^{LL} +
A^{LR} )^{f}_{ij} \; , \vspace{3mm} \\
g_{f R}^{ij} = \; ( A^{RR} + A^{RL} )^{f}_{ij} \; ,
\;\;\;\;\; & {\bar g}_{f R}^{ij} = ( A^{RR} - A^{RL} )^{f}_{ij}
\; ,
\end{array}
\end{eqnarray}
with
\begin{eqnarray}
\label{A relations SM 1}
\begin{array}{ll}
( A^{LL} )^{f}_{ij} = \;
\mid\varepsilon_{L}^{C}\mid^{2}   \; U_{ei}^{*} U_{ej}
 \ \delta_{fe} + \frac{\varrho}{2} \, \varepsilon_{L}^{N \nu} \,
\varepsilon_{L}^{N f}  \, \delta_{ij} \; ,
& (A^{LR} )^{f}_{ij} = \frac{\varrho}{2}
\, \varepsilon_{L}^{N \nu} \varepsilon_{R}^{N f} \, \delta_{ij}
\; , \vspace{3mm} \\
( A^{RR} )^{f}_{ij} =  \; \mid\delta_{R}^{C}\mid^{2}
  \; U_{ei}^{*} U_{ej} \ \delta_{fe} +
\frac{\varrho}{2} \, \delta_{R}^{N \nu} \, \varepsilon_{R}^{N
f}  \, \Omega^R_{ij} \; ,
& (A^{RL} )^{f}_{ij} = \frac{\varrho}{2} \,
\delta_{R}^{N \nu} \varepsilon_{L}^{N f} \, \Omega^R_{ij} \; ,
\end{array}
\end{eqnarray}
where $\varrho = \frac{M_{W}^{2}}{M_{Z}^{2} \cos^{2} \theta_{W}} \simeq 1 $.
For comparison, within the $\nu SM$ only
\begin{eqnarray}
\label{g-c relations SM}
g_{f V}^{ij} = - \bar {g}_{f A}^{ij}  \; &  {\rm and}  & \;
\bar{g}_{f V}^{ij} = - {g}_{f A}^{ij} \;
\end{eqnarray}
are different from zero, reproducing the well-known results for LH
neutrino interactions.
The form of the Hamiltonian in Eq. (\ref{H-eff-sum})
\cite{Wolfenstein_Mikheev_Smirnov,Bergmann 1999}
is especially useful for our purposes.
As neutrinos and background fermions $f= e, p, n$ are placed in
two separated factors, it is straightforward to derive
the coherent neutrino scattering in matter summing over all
fermions and averaging over their properties \cite{Kim_Pevsner}.
This is what is done when going from Eq. (\ref{H-eff-sum}) to
Eq. (\ref{H-eff-nu-copy}) \cite{Bekman 2002}
\begin{eqnarray}
\label{definition}
\begin{array}{l}
(z_{V}^{\ \mu})_{ij} + (z_{V}^{*\ \mu})_{ji} =
(Z_{V}^{\ \mu})_{ij} = \frac{G_{F}}{\sqrt{2}} \sum_{f}
N_{f} (g_{fV}^{ij} \langle\frac{p^{\mu}_{f}}{E_{f}}\rangle + m_{f}
\overline{g}_{fV}^{ij}\langle\frac{s_{f}^{\mu}}{E_{f}}\rangle) \; ,
\vspace{3mm} \\
(z_{A}^{\ \mu})_{ij} + (z_{A}^{*\ \mu})_{ji} =
(Z_{A}^{\ \mu})_{ij} = \frac{G_{F}}{\sqrt{2}} \sum_{f}
N_{f}(\overline{g}_{fA}^{ij}
\langle\frac{p^{\mu}_{f}}{E_{f}}\rangle + m_{f}
g_{fA}^{ij}\langle\frac{s_{f}^{\mu}}{E_{f}}\rangle) \; ,
\end{array}
\end{eqnarray}
where $N_{f}$ with $f = e, p, n$ stand for the number of
fermions $f$ per unit volume,
and $\langle\frac{p^{\mu}_{f}}{E_{f}}\rangle$
and $\langle\frac{s_{f}^{\mu}}{E_{f}}\rangle$
are the properly normalised averages over the corresponding
fermion distributions of the momentum and the spin, respectively.

\subsection{The effective, coherent  neutrino interaction in
matter}

The neutrino background coherent scattering is now
easy to calculate from the effective four-fermion Hamiltonian.
Propagation of the light, relativistic neutrinos in matter is
governed by an evolution equation of the Schr\"odinger type for
each spinor helicity for this is conserved
(see e.g. \cite{Bekman 2002,sev}):
\begin{eqnarray}
\label{Schrod. Equation} i\frac{d}{dt}\ \Psi_{i}^{
\nu}(\overrightarrow{p},t)\ =\ \sum_{j}\ {\cal{H}}^{ \nu}_{ij}\
\Psi_{j}^{ \nu}(\overrightarrow{p},t) \; ,
\end{eqnarray}
where $\Psi_{i}^{ \nu}(\overrightarrow{p},t)$ is the wave function
for the neutrino (antineutrino) eigenstate of mass $m_{i}$,
momentum ${\overrightarrow{p}}$ and helicity
$\frac{\lambda}{2} = - \frac{1}{2} \, ( + \frac{1}{2})$.
As usual \cite{Giunti Kim,Kim_Pevsner},
we assume that all particles have the same momentum but different
energies $E_{i}=\sqrt{\overrightarrow{p}^{2} + m_{i}^{2}}$.
Then, the effective Hamitonian in Eq. (\ref{Schrod. Equation})
reads
\begin{eqnarray}
\label{Effective Ham}
{\cal{H}}_{ij}^{ \nu} \  =\  (p + \frac{m_{i}^{2}}{2 p})\
\delta_{ij}\  +\  {\cal{H}}_{ij} \; ,
\end{eqnarray}
where ${\cal{H}}_{ij}$ describes the coherent neutrino
scattering inside matter,
thus depending on the properties of the medium which is
characterised by the fermion contents and the corresponding
polarization, momentum and charge distributions. As it is
apparent from Eqs. (\ref{M-i-k-Dirac-rel 1}) and
(\ref{definition}), ${\cal{H}}_{ij}$
is also different for Dirac and Majorana neutrinos. In the case of
a medium unpolarized, isotropic and neutral ($N_{e} = N_{p}$),
the Hamiltonian for Dirac neutrinos with helicity
$\frac{\lambda}{2}= - \frac{1}{2}$ is
\begin{eqnarray}
\label{Dirac(-1)}
{\cal{H}}^{D}_{ij}\ (\lambda=-1) = \sqrt{2} \,
G_{F} \left[ N_{e} (g^{ij}_{e L} + g^{ij}_{p L}) + N_{n}
g^{ij}_{n L} \right] \; ,
\end{eqnarray}
whereas for Majorana neutrinos of the same helicity it
reads
\begin{eqnarray}
\label{Majorana(-1)}
{\cal{H}}^{M}_{ij}\ (\lambda=-1) = \sqrt{2}
\, G_{F} \left[ N_{e} (g^{ij}_{e L} - g^{* ij}_{e R} + g^{ij}_{p
L} - g^{* ij}_{p R}) + N_{n} (g^{ij}_{n L} - g^{* ij}_{n R})
\right] \; .
\end{eqnarray}
Thus, Dirac and Majorana neutrinos propagate differently in
matter in the presence of extra RH interactions, ${g_{f R} {\neq 0}}$.
Their evolution depends on the matter densities $N_{f}$
and the neutrino couplings to electrons, protons and
neutrons $g_{f}$. These are related
to the $u$ and $d$ quark couplings in the usual way
\begin{eqnarray}
\label{M-M (-1)-sum}
\begin{array}{l}
g^{ij}_{p L}  = 2 g^{ij}_{u L} + g^{ij}_{d L}
\; , \;\;\;\; g^{ij}_{p R}  = 2 g^{ij}_{u R} + g^{ij}_{d R}
\; , \vspace{3mm} \\
g^{ij}_{n L}  = g^{ij}_{u L} + 2 g^{ij}_{d L} \; , \;\;\;\;
g^{ij}_{n R}  = g^{ij}_{u R} + 2 g^{ij}_{d R} \; .
\end{array}
\end{eqnarray}
The appropriate Hamitonians for Dirac antineutrinos and
Majorana neutrinos of positive helicity follow from the well-known
relations \cite{Bekman 2002}
\begin{eqnarray}
\label{H-D-D-bar H-M-M-bar}
{\cal{H}}^{D}_{{\overline i}{\overline j}} (\lambda) \
=\ -\ [
 {\cal{H}}^{D}_{ij} ( - \lambda) ]^{\; *} \; , \;\;\;\;
{\cal{H}}^{M}_{ij}(\lambda) = -
[{\cal{H}}^{M}_{ij} ( - \lambda) ]^{\; *} \; ,
\end{eqnarray}
which can be also deduced from Eq. (\ref{M-i-k-Dirac-rel 1}).
We should remember that these relations are valid in the absence of
scalar, pseudoscalar and tensor interactions, for relativistic
neutrinos, and in the mass eigenstate basis. In the flavour basis
these relations are in general not satisfied.

\section{Propagation of Dirac and Majorana neutrinos in matter}
\label{Propagation}

In order to estimate the difference between Dirac and
Majorana neutrino oscillations we shall consider the simple case
discussed in the previous Section of an unpolarized, isotropic and
neutral medium with constant fermion densities $N_{f}$.
Then, the neutrino evolution equation in Eq. (\ref{Schrod. Equation})
can be solved analytically.
After substracting the non-important common diagonal pieces,
the Hamiltonians (\ref{Dirac(-1)}) and (\ref{Majorana(-1)}) can be
written in the flavour eigenstate basis
\begin{eqnarray}
\label{Dirac2} {\cal{H}}_{\alpha\beta}^{D}\ (\lambda=-1) = \sqrt{2}
\, G_{F}  N_{e}\   \delta_{e\alpha}\   \delta_{e\beta} \
\end{eqnarray}
and
\begin{eqnarray}
\label{Majorana2}
\begin{array}{rl}
{\cal{H}}_{\alpha\beta}^{M} (\lambda=-1) =
&\sqrt{2} \, G_{F} \biggl\{ \, N_{e}\   \delta_{e\alpha}\
\delta_{e\beta}
 - N_{e} [ \,
|\delta_{R}^{C}|^{2}\ \delta_{e\alpha}\
 \delta_{e\beta} \vspace{2mm} \\
&+\ \frac{1}{2} \ \delta_{R}^{N\nu *}\
 (\delta_{L}^{e*} + \delta_{R}^{e*} + 2 \delta_{L}^{u*} +
2 \delta_{R}^{u*} +  \delta_{L}^{d*} +
 \delta_{R}^{d*}) \, \Omega^{R *}_{\alpha\beta} \, ] \vspace{2mm} \\
&-\ N_{n}\
\frac{1}{2}\ \delta_{R}^{N\nu *}\ (-1 + 2 \delta_{L}^{d*}
+ 2 \delta_{R}^{d*} + \delta_{L}^{u*} +
 \delta_{R}^{u*}) \, \Omega^{R *}_{\alpha\beta} \, \biggr\} \ ,
\end{array}
\end{eqnarray}
respectively, with
\begin{eqnarray}
\label{Matrix Omega} \Omega^R_{\alpha\beta} = \sum_{i,j}U_{\alpha
i}\ \Omega^R_{i j}\  U_{\beta j}^{*}\ .
\end{eqnarray}
The new RH interactions do not contribute to the Dirac effective
Hamiltonian in Eq. (\ref{Dirac2}), which is the same as in the
$\nu SM $, but they do contribute to the Majorana one in Eq.
(\ref{Majorana2}). Neglecting quadratic terms in the new small
parameters, this reduces to
\begin{eqnarray}
\label{Majorana3}
{\cal{H}}_{\alpha\beta}^{M}(\lambda=-1) =
\sqrt{2} \, G_{F}  N_{e} \{\   \delta_{e\alpha}\   \delta_{e\beta}
+
  \frac{N_{n}}{N_{e}} \frac{1}{2}
 \delta_{R}^{N\nu *} \Omega_{\alpha\beta}^{R *}\ \}\ .
\end{eqnarray}
Thus, the difference between Dirac and Majorana
neutrino oscillations are triggered by the extra term proportional
to the neutron to electron density ratio, to
$\delta_{R}^{N\nu *}$ and to the
$\Omega_{\alpha\beta}^{R *}$ matrix elements.
Then, to observe any difference, $\Omega_{\alpha\beta}^{R}$ can not
be diagonal.
It is also worth to emphasize that in contrast with other NP effects,
which are quadratic in the (small) new parameters,
this difference is linear in the (small) strength of the extra
RH neutral interactions.

The oscillation probabilities can be found diagonalising the
Hamiltonians in Eqs. (\ref{Dirac2}) and (\ref{Majorana3}) together
with the neutrino (species) dependent kinematical term in Eq.
(\ref{Effective Ham}) \cite{Kim_Pevsner, Bekman 2002}
\begin{eqnarray}
\label{mdiagonal} {\cal{H}}^{\nu}\  =\  \frac{1}{2 E_{\nu}}\
{W} \ \{ {\sf diag}\ ( \tilde{m}_i^2 ) \} \ {W}^{\dagger} \; ,
\end{eqnarray}
where ${W}$ is the diagonalising (unitary) matrix defined by the
eigenvectors of $2 E_{\nu} {\cal{H}^{\nu}}$ and
$\tilde{m}_i^2$ are the corresponding (real) eigenvalues.
We are now ready to calculate the transition amplitude from
the initial neutrino (production) state $|\psi(0)>$ after
travelling a distance $L$ to some other final neutrino
(detection) state $|\varphi(0)>$,
\begin{eqnarray*}
<\varphi(0)|\psi(L=t)> .
\end{eqnarray*}

When considering new neutrino interactions these can also affect
their production and detection (see e.g. \cite{Gonzalez}).
In general, to calculate the impact of NP on the full process, the
modification of the initial and final states should be also taken
into account. But here we are interested in the difference between
Dirac and Majorana neutrino propagation, and the effects which
modify in the same way both type of neutrinos can be ignored.
Therefore, we can assume that neutrinos are produced and detected
in the flavour states $| {\nu}_{\alpha} >$ and $| {\nu}_{\beta} >$,
respectively. Then the transition amplitude
${\nu}_{\alpha}\rightarrow{\nu}_{\beta}$
can be written
\begin{eqnarray}
\label{amplitude}
A_{\alpha \to \beta} (L)\  =\
<{\nu}_{\beta}(0)|{\nu}_{\alpha}(L=t)>\  =\  (W\  \{ {\sf diag}\
(\exp\ [ - i \frac{{\tilde{m}}_{i}^{2}}{2 E_{\nu}} L ] )\}
\ W^{\dag})_{\beta\alpha} ,
\end{eqnarray}
and the transition probability \cite{Gluza_Zralek}
\begin{eqnarray}
\label{P alfa w beta}
P_{\alpha \to \beta} (L) &=&
{\delta}_{\alpha \beta} - 4 \sum\limits_{a>b} {R}^{ab}_{\alpha
\beta} \sin^2{\Delta_{ab}} + 8 {I}^{21}_{\alpha \beta}
\sin{\Delta_{21}} \sin{\Delta_{31}} \sin{\Delta_{32}} \ ,
\end{eqnarray}
with
\begin{eqnarray}
\label{R,I,W}
{R}^{ab}_{\alpha \beta} = Re \left[ W_{\alpha a}^{\ast}
W_{\beta a} W_{\alpha b} W_{\beta b}^{\ast}  \right] \; ,
\;\;\; {I}^{21}_{\alpha \beta} = Im \left[ W_{\alpha 2}^{\ast}
W_{\beta 2} W_{\alpha 1} W_{\beta 1}^{\ast}   \right] \; ,
\end{eqnarray}
and
\begin{eqnarray}
\Delta_{ab} = 1.27 \frac{(\tilde{m}_a^2 -
\tilde{m}_b^2) [{\rm eV}^2] \,  L [{\rm km}] }{E_{\nu} [{\rm GeV}] } \, .
\end{eqnarray}
Using Eq. (\ref{H-D-D-bar H-M-M-bar}) we can obtain the transition
probability for Dirac antineutrinos and Majorana neutrinos with
$\lambda = + 1$ from Eq. (\ref{P alfa w beta}) with the appropriate
replacements
\begin{eqnarray}
P_{{\overline{\alpha}} \to {\overline{\beta}}} (L) = P_{\alpha
\to \beta} (L; {W} \to {W}^\ast, G_F \to -G_F ).
\end{eqnarray}

\section{Numerical results and experimental bounds}
\label{bounds}

In this Section we quantify how large can the
difference between the Dirac and Majorana neutrino oscillation
probabilities be in the presence of new RH neutral interactions.
We shall assume that $\delta_{R}^{N \, \nu}$ and the matrices
$U^{*}$ and $\Omega^{R}$  in Eqs. (\ref{L-int-CC-general}) and
(\ref{L-int-NC-general}) are real, neglecting possible CP
violating phases because they do not change our results in any
substantial way. For the calculations shown in the Figures below
we use the first order expressions in $\delta_{R}^{N  \nu}
\Omega^{R}$ for the Dirac and Majorana effective Hamiltonians in
Eqs. (\ref{Dirac2}) and (\ref{Majorana3}), respectively.
We shall take
$\delta_{R}^{N  \nu}$ not bigger than 0.01 \cite{Sobkow,
Gonzalez_Maltoni_ogr_param}, and parameterise $\Omega^{R}$ as
follows
\begin{eqnarray}
\label{OmegaRij}
\Omega^{R}_{\alpha \beta} = \left( \matrix{ 1
& {\eta} & {\eta^{2}}  \cr {\eta} & 1 - \chi & {\eta} \cr
                        {\eta^{2}} & {\eta} & 1 - \omega} \right) \; .
\end{eqnarray}
We have checked that the difference of the Majorana and Dirac
neutrino transition probabilities $\Delta P = P^{M} - P^{D}$
depends litlle on the diagonal elements $(\chi, \, \omega)$ for
any choice of sign. In contrast the dependence on the off-diagonal
entry $\eta$ is linear
\footnote{A quantitative discussion of the effects of flavour
diagonal and off-diagonal non-standard Hamiltonian contributions
to neutrino oscillations can be found in Ref. \cite{Winter}.}.
We choose this particular $\eta$ parameterisation
of $\Omega^{R}$ only for easy comparison of the potential of the
different channels to discriminate between Dirac and Majorana
neutrinos. The oscillation probabilities also depend on the
medium. We shall concentrate on the neutrino propagation inside
Earth, allowing for variations of the travelling distance $L$ and
of the neutrino energy $E_{\nu}$. The matter density changes along
the Earth radius, and so the electron (proton) and neutron
densities in Eqs. (\ref{Dirac2}) and (\ref{Majorana3}). Our
calculations are performed for travelling distances $L = 1000$,
$6500$ and $13000$ km using along these paths the mean values of
the matter densities, which we estimate to be equal to $3$, $4$
and $7$ ${\rm g/cm^{3}}$, respectively. As we are only interested
in illustrating the main trends, we shall not consider realistic
profiles of the Earth matter density. We also need to know $U$.
Using the solar, KamLAND and SK+K2K+CHOOZ data, the values of
the five standard oscillation parameters
$\theta_{12}$,
$\theta_{13}$,
$\theta_{23}$,
$\delta m^{2}_{21} \equiv \delta m^{2}_{sol}$ and
$\delta m^2_{32} \equiv \delta m^{2}_{atm}$
can be determined. At 95\% C.L. \cite{Fogli}
\begin{eqnarray}
\label{Data1}
\sin^{2}\theta_{12} &=& 0.314\ (1^{+0.18}_{-0.15}) \; , \nonumber \\
\delta m^{2}_{21} &=&
7.92\ (1\pm0.09) \times 10^{-5} \; {\rm eV^{2}} \; ,  \nonumber \\
\sin^{2}\theta_{23} &=& 0.44\ (1^{+0.41}_{-0.22}) \; , \\
\delta m^{2}_{32} &=& 2.4\ (1^{+0.21}_{-0.26}) \times 10^{-3} \;
{\rm eV^{2}} \; , \nonumber  \\
\sin^{2}\theta_{13} &=&
0.9\ ^{+2.3}_{-0.9} \times 10^{-2} \; . \nonumber
\end{eqnarray}

The numerical results are presented in Figs. $1 - 5$. We plot the
values of the transition probabilities for Dirac and Majorana
neutrinos. Their difference $\Delta P$ is of
few per cent. Figs. 1 (a) $- 1$ (f) prove that the transition
probabilities also depend on the neutrino flavour. For $\nu_{\mu}
\rightarrow \nu_{e}$ the effect first increases, and then it seems
to weaken again with increasing $L$. But for the $\nu_{\mu}
\rightarrow \nu_{\tau}$ channel the probability
difference $\Delta P$ keeps increasing with increasing
distance $L$ up to the last analyzed value $L = 13000$ km.
\begin{figure}[top]
\begin{center}
\subfigure{\includegraphics[angle=0,width=70mm]{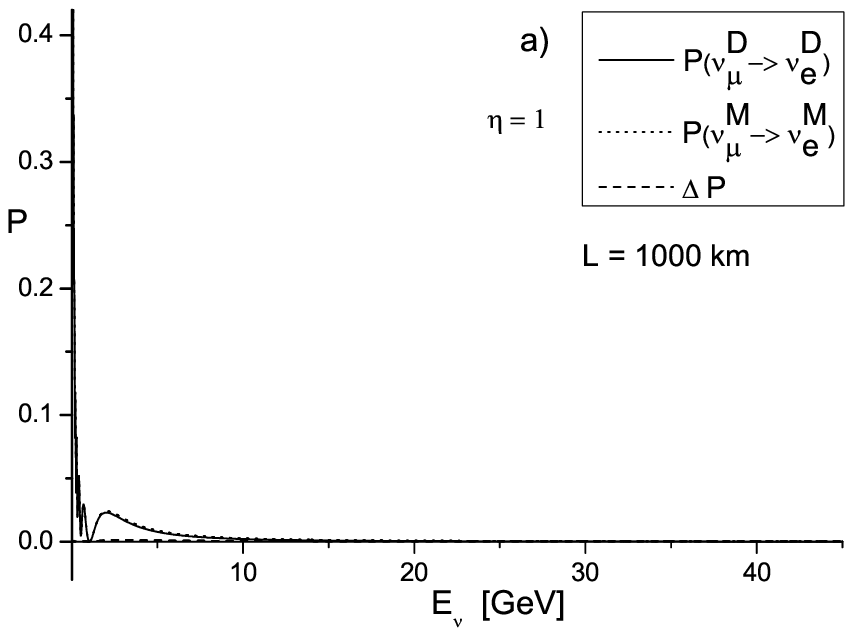}}
~~~~~~~~~~
\subfigure{\includegraphics[angle=0,width=70mm]{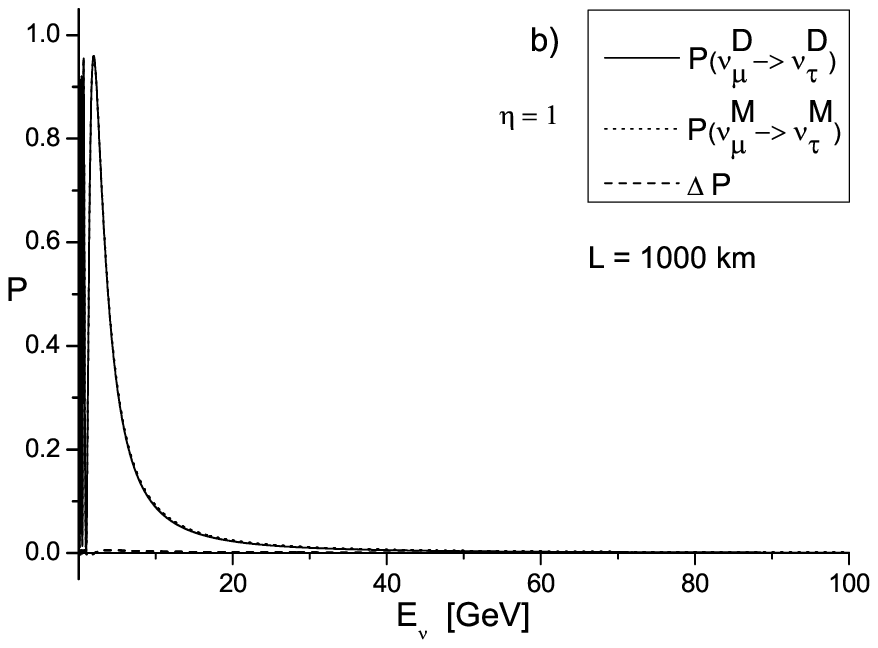}}
\subfigure{\includegraphics[angle=0,width=70mm]{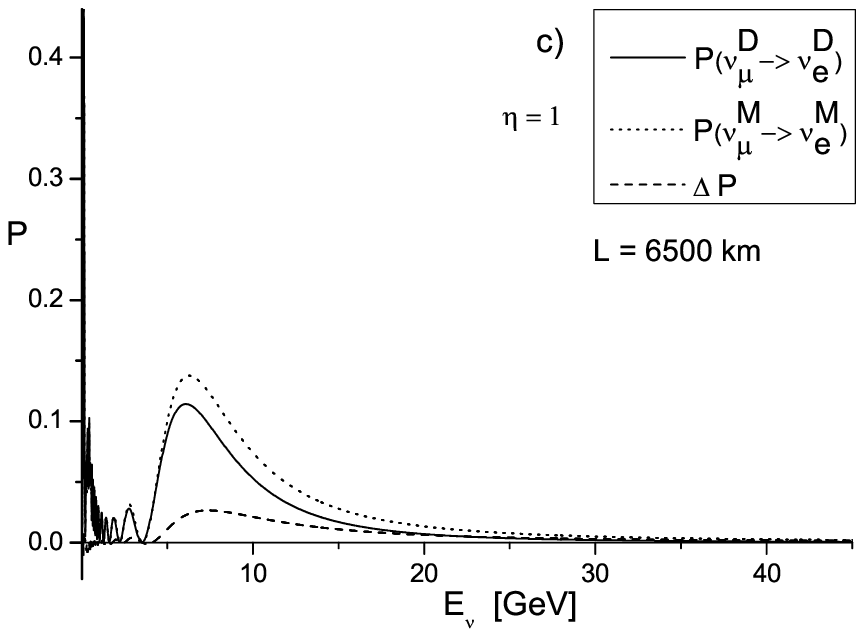}}
~~~~~~~~~~
\subfigure{\includegraphics[angle=0,width=70mm]{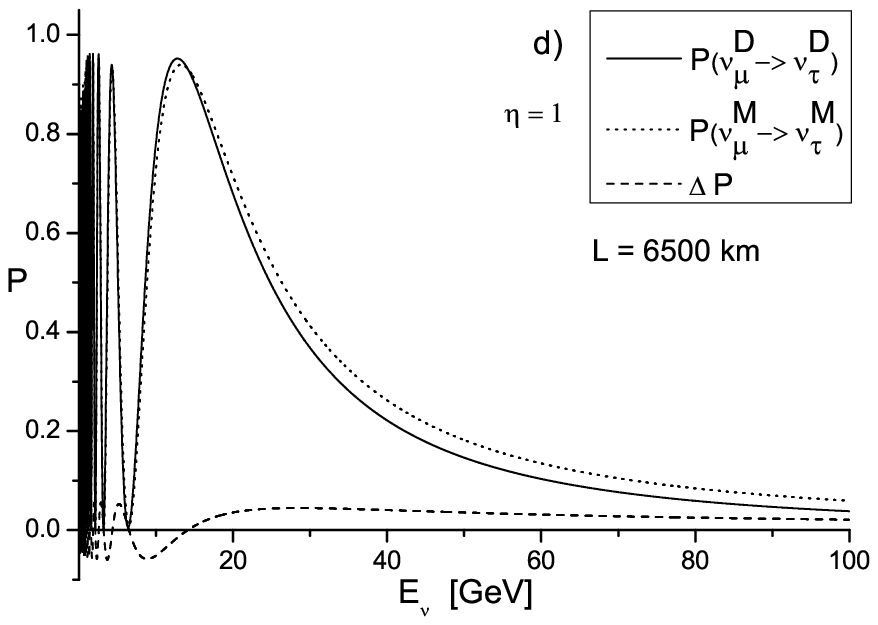}}
\subfigure{\includegraphics[angle=0,width=70mm]{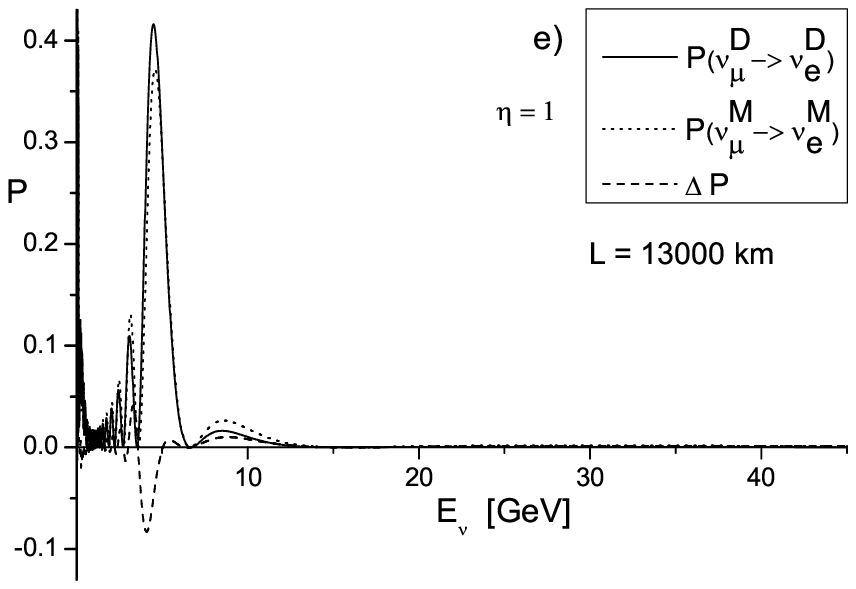}}
~~~~~~~~~~
\subfigure{\includegraphics[angle=0,width=70mm]{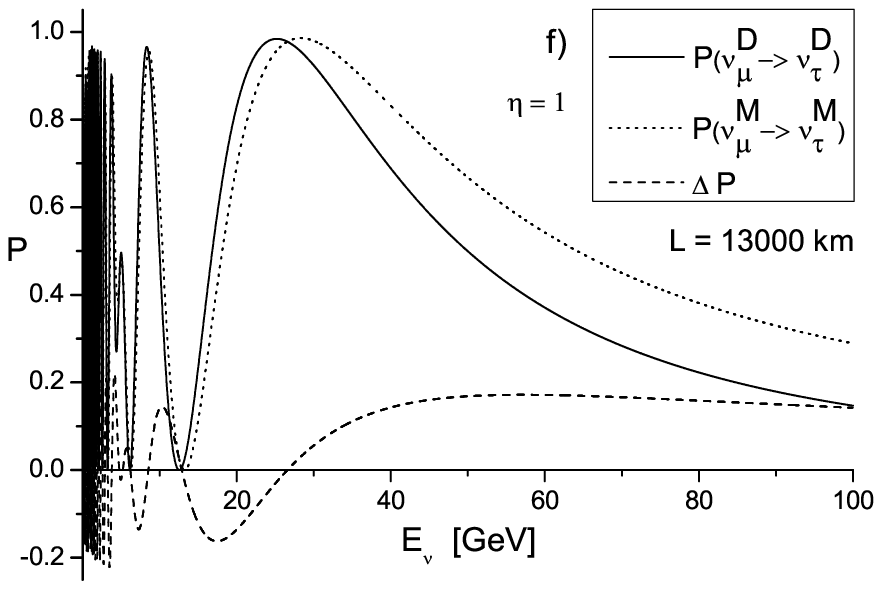}}
\caption{Transition probabilities for Majorana and
Dirac neutrinos and their difference $\Delta P$ as a function
of the neutrino energy $E_{\nu}$ GeV and $\eta = 1$. {\bf From left
to right and top to bottom}:
(a)~$\nu_{\mu} \rightarrow \nu_{e}$,  $L = 1000$ km.
(b)~$\nu_{\mu} \rightarrow \nu_{\tau}$,  $L = 1000$ km.
(c)~$\nu_{\mu} \rightarrow \nu_{e}$,  $L = 6500$ km.
(d)~$\nu_{\mu} \rightarrow \nu_{\tau}$,  $L = 6500$ km.
(e)~$\nu_{\mu} \rightarrow \nu_{e}$,  $L = 13000$ km.
(f)~$\nu_{\mu} \rightarrow \nu_{\tau}$,  $L = 13000$ km.}
\end{center}
\end{figure}
As observed in these Figures the largest effects manifest in the
$\nu_{\mu} \rightarrow \nu_{\tau}$ transitions when $L = 13000$ km
(approximate diameter of the Earth). In this case the largest
difference $\Delta P$ between Majorana and Dirac neutrino
transition probabilities increases with energy, and for $\eta = 1$
reaches its maximum at $E_{\nu} \approx 57$ GeV with a long
(experimentally attractive) plateau at higher energies.
It varies approximately between 0.15 and 0.17, standing for
an effect on the value of $\Delta P/P^{D}$ equal to 34 \% at 50
GeV, increasing up to 84 \% at 90 GeV. It can be seen
too that for small neutrino energies the transition probability
difference can be significant as well. However, this is of no
practical use because of the high frequency of the probability
variation, which after averaging over the energy bin smooths
the signal to zero. Figs. 2 (a) $-$ (b) show the $\eta$ dependence. The
effect weakens with the decrease of $\eta$ for it is proportional
to $\delta_{R}^{N \, \nu} \eta$. Indeed, for $\eta = 1/5$ the
largest difference between the transition probabilities is reached
at $E_{\nu} \approx 53$ GeV, with a wide plateau too but with a
similar value of $\Delta P \approx 0.035$ (see Fig. 2 (a)). To make it
apparent we draw in Fig. 2 (b) the $\eta$ dependance of $\Delta P$ in
the vecinity of the maximum. Thus, for $L = 13000$ km and $E_{\nu}
= 57$ GeV we vary $\eta$ up to $1.0$. As expected the effect
decreases linearly with $\eta$ vanishing for $\eta = 0$.
\begin{figure}[top]
\begin{center}
\subfigure{\includegraphics[angle=0,width=70mm]{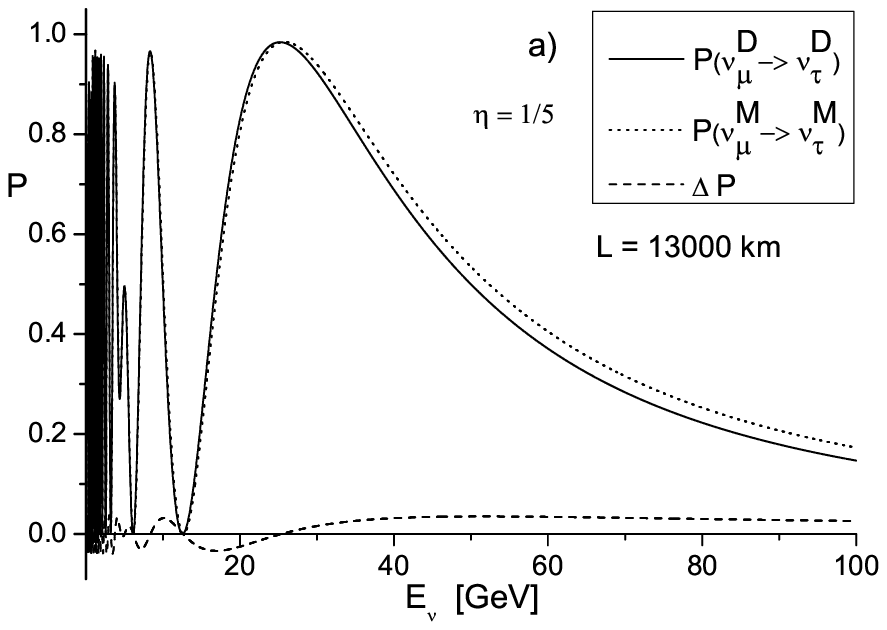}}
~~~~~~~~~~
\subfigure{\includegraphics[angle=0,width=70mm]{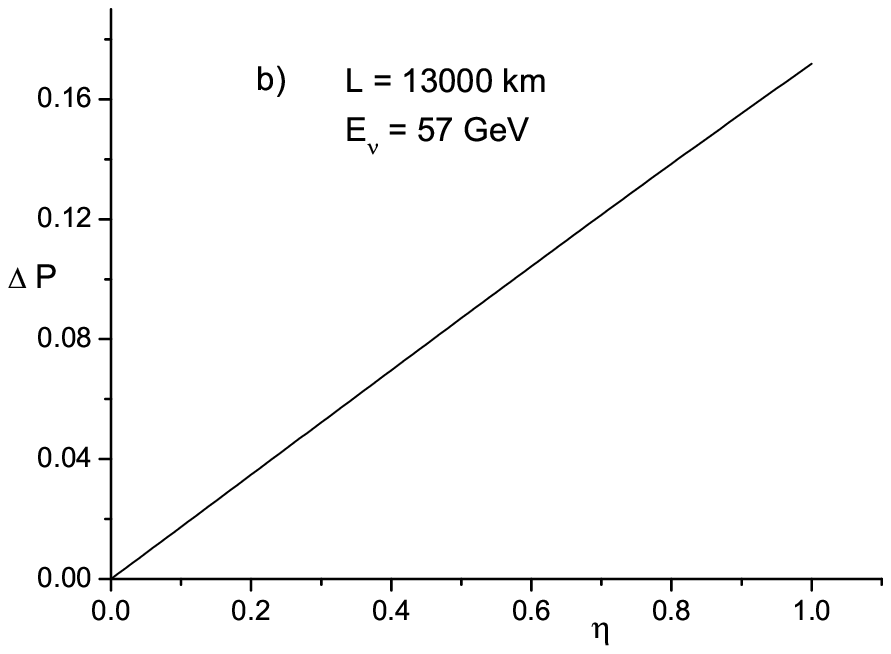}}
\caption{{\bf Left}: (a) Transition probabilities for
Dirac and Majorana neutrinos and their difference $\Delta
P$ as a function of the neutrino energy $E_{\nu}$ in GeV for $L
= 13000$ km and for the process $\nu_{\mu} \rightarrow
\nu_{\tau}$, with $\eta = 1/5$. {\bf Right}: (b) Transition probability
difference $\Delta P$ for the process $\nu_{\mu} \rightarrow
\nu_{\tau}$ as a function of $\eta$ for a neutrino energy
$E_{\nu} = 57$ GeV and $L = 13000$ km.}
\end{center}
\end{figure}
Finally, in order to visualize the two-dimensional dependance of
$\Delta P$ on $L$ and $E_{\nu}$ we include Fig. 3.
For $E_{\nu} \leq 20$ GeV, $\Delta P$ changes very rapidly as a
function of the baseline $L$. For larger energies the variation is
slower.
\begin{figure}[top]
\begin{center}
\subfigure{\includegraphics[angle=0,width=120mm]{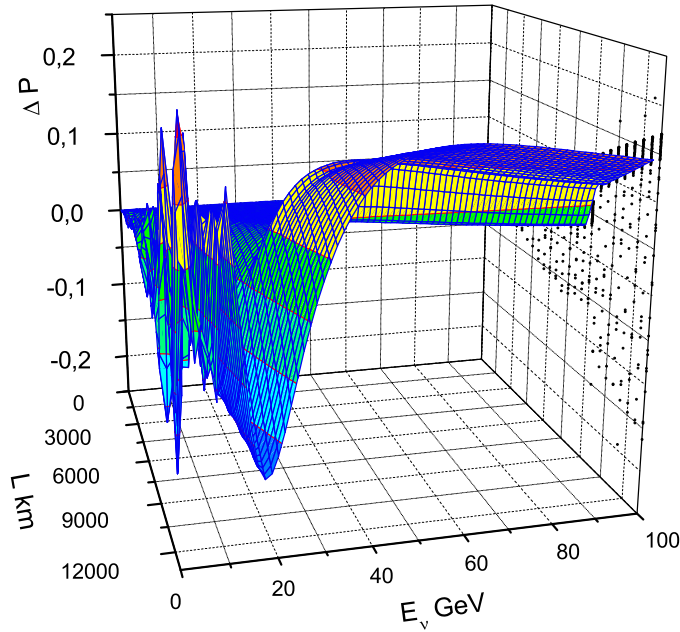}}
\caption{Transition probability difference $\Delta P$ as
a function of the neutrino energy $E_{\nu}$ in GeV and the distance
$L$ in km for the process $\nu_{\mu} \rightarrow \nu_{\tau}$
with $\eta = 1$.
The calculations are performed for matter densities equal
to $3$, $4$ and $7$ $[g/cm^{3}]$ and travelling distances $L$
between $0-4500$, $5000-9000$ and $9500-13000$ km, respectively.
The projections on the $L$-$\Delta P$ plane are drawn with dots.}
\end{center}
\end{figure}

The question arises whether current experimental errors
\cite{Gonzales_Maltoni, sno, Fogli} are not too weak to prevent
the transition probabilities for Dirac and Majorana neutrinos
from overlapping.
We show in Fig. 4 (a) the $\nu_{\mu}\rightarrow \nu_{\tau}$
transition probabilities for present values of the oscillation
parameters (Eq. (\ref{Data1})). The bands are obtained varying
the oscillation parameters within their 95\% C.L. limits.
Indeed, they are too wide to be able to distinguish
between Dirac and Majorana neutrinos. However, future
experiments will provide more precise measurements.
Hence, we plot in Fig. 4 (b) the bands for the same central
values of the oscillation parameters but assuming that the
errors are reduced by a factor of 5.
In this case the Majorana (upper) and Dirac (lower) bands
separate, allowing in principle to distinguish between both
types of neutrinos.
\begin{figure}[top]
\begin{center}
\subfigure{\includegraphics[angle=0,width=75mm]{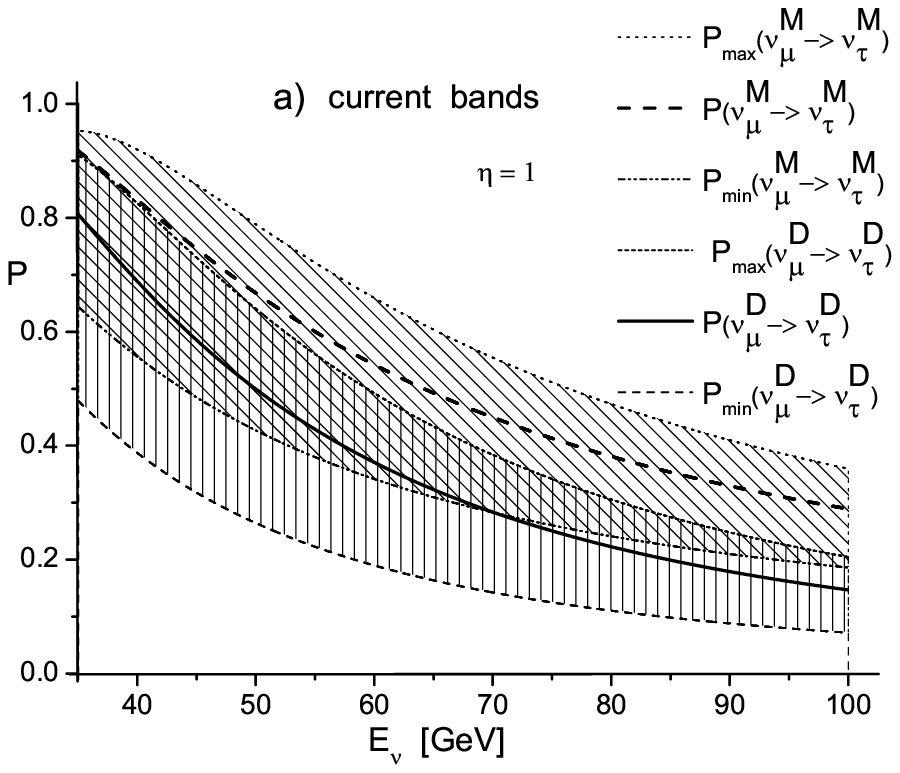}}
~~~~~~~~~~
\subfigure{\includegraphics[angle=0,width=70mm]{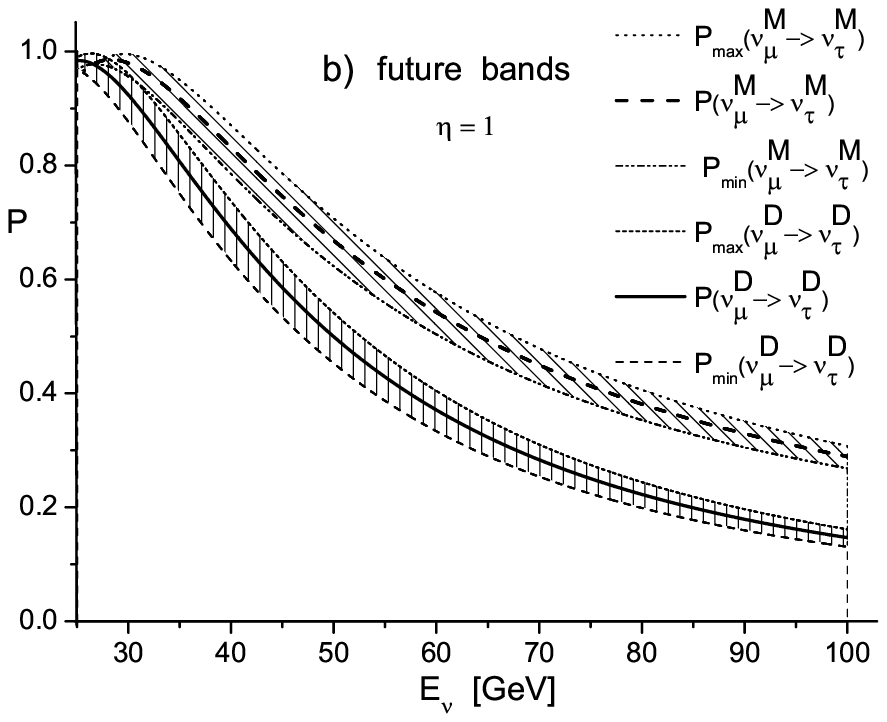}}
\caption{ Upper and lower bands corresponding to transition
probabilities for Dirac and Majorana neutrino, respectively, as a
function of the neutrino energy $E_{\nu}$ in GeV for $L = 13000$
km and the process $\nu_{\mu} \rightarrow \nu_{\tau}$ with $\eta = 1$.
{\bf Left}: (a) For current 95 \% confidence intervals of the
oscillation parameters \cite{Fogli}. {\bf Right}: (b) For future 95 \%
confidence intervals of the oscillation parameters (assuming
a statistics $\approx 25$ times larger than nowadays).}
\label{difference}
\end{center}
\end{figure}
Obviously varying $\eta$ (and the form of $\Omega$) one can
make the effect much smaller and unobservable.

\begin{figure}[top]
\begin{center}
\subfigure{\includegraphics[angle=0,width=100mm]
{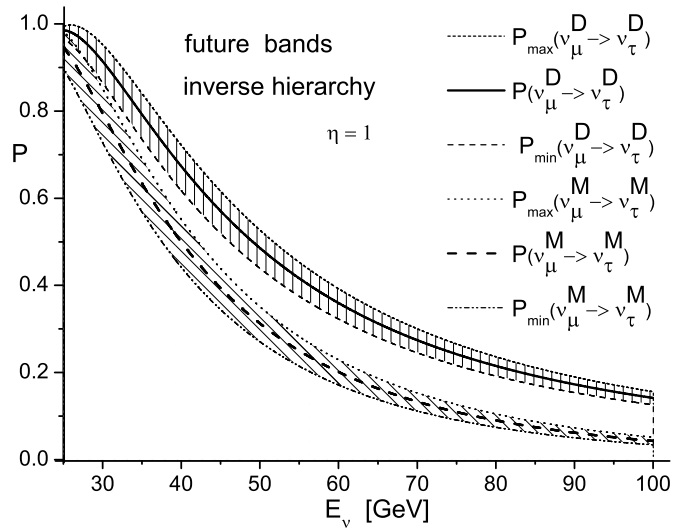}}
\caption{
Transition probabilities for Dirac (upper band) and
Majorana (lower band) neutrinos in the inverse mass
scheme
(assuming a statistics $\approx$ 25 times larger than nowadays).
There is almost no difference between the Dirac neutrino bands
for the inverted and normal schemes (see Fig. \ref{difference} (b)).
}
\label{scheme}
\end{center}
\end{figure}

This encourages to search for the neutrino character at future
experiments. One must be aware, however, that not only the
experimental errors of the standard oscillation parameters must be
significantly reduced but they must be determined independently to
avoid new confusion \cite{Confusion}. A relevant related comment
is that the observation of deviations from the SM predictions, as
those drawn for Majorana neutrinos in the presence of RH neutral
currents in Fig. \ref{difference}, can have their origin in the NP
previously discussed or be a manifestation of the inverted
character of the neutrino mass hierarchy. The distances and
energies relevant in our case are similar to those sensitive to
atmospheric neutrino oscillations, and then to $\delta m^2_{32}$.
Its sign, which is unknown at present and may be positive (normal)
or negative (inverted scheme), can not be fixed in the $\nu _\mu
\rightarrow \nu _\tau$ channel within the $\nu SM$ \footnote{If
$\theta _{13}$ is sizeable, the sign of $\delta m^2_{atm}$ will be
possibly established in $\nu _\mu \rightarrow \nu _e$ oscillation
experiments \cite{theta13}, even if there are no new interactions.
In our case the relevant channel is $\nu _\mu \rightarrow \nu
_\tau$ because it is where the effect of new RH neutral
interactions is larger.}. Therefore in Fig. \ref{scheme} we plot
the same transition probabilities as in Fig. \ref{difference} (b)
but for the inverted scheme.
The central lines and future bands for Dirac neutrinos,
which give the same predictions as the $\nu SM$,
almost coincide for both sign assignments.
In contrast, the band for Majorana neutrinos moves below
(above) the Dirac neutrino band for the inverted (normal)
hierarchy, reversing the sign of $\Delta P$ but maintaining
its size to a large extent.
The same displacement is obtained changing the sign
of $\eta$. This is well-known
\cite{Kayser}. The sign of
$\delta m^2_{atm}$ can be fixed if there is other
contribution to the interaction potential energy with
a well-defined sign, with which the standard
contribution can interfere. In our case the extra
piece is provided by the new RH interactions. If in
the future any of the two schemes is established,
the observation of the Majorana character of the
light neutrinos in the process in Figs. \ref{difference}
(b) and \ref{scheme} would also allow for the determination
of the sign of $\eta$.

\section{Conclusions and models}

Within the $\nu SM$ oscillations of relativistic
neutrinos do not differentiate between Dirac and Majorana
neutrinos. This is not in general the case beyond the $\nu SM$,
and Dirac and Majorana neutrinos can be differentiated by
how they propagate in matter.
Indeed, extended models with new RH currents or tensor
terms do distinguish between both types of neutrinos.
However, the spin-flip tensor
transitions require a polarised medium to
manifest \cite{Bergmann 1999}. Moreover, the
production and detection of (anti)neutrinos of the wrong helicity
are strongly suppressed.
Hence, we concentrated on the analysis of additional RH currents
which allow for unsuppressed spin non-flip transitions.
This NP does not
modify the oscillation of Dirac neutrinos in matter for
we can assign to them a definite lepton number,
but it does modify the propagation of
Majorana neutrinos for they participate of both types (LH and RH)
of currents. The largest effect, which is linear in the new small
parameters, is associated to extra RH neutral currents. However,
it is needed a non-trivial flavour structure to differentiate
Dirac from Majorana neutrino oscillations in matter. Both LH and
RH neutral current interactions can not be simultaneously
diagonal.

Numerical estimates show that the largest differences between
Majorana and Dirac neutrino oscillations manifest in the
$\nu_{\mu}\rightarrow \nu_{\tau}$ channel. This difference between
both probabilities can be as large as $\sim 0.16$
for the normal as well as for the inverted scheme, but with
opposite sign.
Unfortunately, this large effect holds for the channel which is
more difficult to measure and for a very large baseline
$L\simeq 13000 \ \rm km$.
Smaller effects are visible for $\nu_{\mu}\rightarrow \nu_{e}$ and
other baselines.

The large effect found is a consequence of the linear, and then
unsuppressed, dependence of the transition probabilities for
Majorana neutrinos on the new RH neutral currents.
This, which is our main observation, relies on two
ingredients for being in practice of some relevance.
The strength of the new flavour violating RH transitions
has to be sufficiently large, and the precision reached in
oscillation experiments sufficiently high. We have not
attempted a global fit to determine the present experimental
limits on the new (flavour changing) couplings but we have
looked for a class of $\nu SM$ extensions which may accommodate
such new terms.
Let us present a simple example of a class of extended gauge
models of possible cosmological interest \cite{zp3}. Consider an
extended gauge group with an extra $U(1)_N$ under which all known
particles transform trivially, and enlarge the matter content with
two new fermion singlets under the $\nu SM$ but with $U(1)_N$
charges $2$ and $-2$ (what makes the model free of gauge
anomalies) and the same lepton number (what forbids direct mass
terms in the Dirac case). This is completed with two extra Higgs
doublets $h'_i$ and an extra Higgs singlet $\phi$
with $U(1)_N$ charges $2, -2$ and $1$, respectively.
The new fermions get their masses through
their very small Yukawa couplings with the $\nu SM$ neutrinos,
once $h'_i$ get non-zero vacuum expectation values (vev) $v'_i$,
also (in principle) much smaller than the $\nu SM$ Higgs vev
$< h >\  = v$.
Whereas the singlet vev $< \phi >\ = x$ is (much) larger than $v$,
and gives a mass to the new gauge boson.
The new RH neutral current term for the light neutrinos
is proportional to the $Z$ boson mixing with the new gauge boson,
which also scales with $v'_i$. The $\nu SM$ is practically
recovered for $< h'_i > \ = 0$. Then, the new phenomenology of this
model depends on those vev. In the absence of Majorana masses
lepton number is conserved and there are one massless (LH) neutrino
and two light Dirac neutrinos. If there are also light neutrino
masses, we have five light Majorana neutrinos. In the first case
the sum in the extra RH piece is over the new RH fermions only,
while in the second one the sums are over all the Majorana neutrinos
\footnote{The LH charged current sum also extends to the
new light neutrinos.}.
One can write down more complicated models with at least
three massive Dirac neutrinos. Nevertheless, the important
question is if the new gauge boson and the extra scalars can
escape detection and at the same time the LH interactions approach
the $\nu SM$ ones, as required by experiment, whereas the new
RH piece has flavour changing couplings large enough, $\sim$
100 times smaller than the standard couplings. The main tension
manifests in the Majorana case because the effective RH
couplings are the product of the $Z$ mixing with the new gauge boson
times the square of the mixing of the standard
neutrinos with the new singlets, and none of them can be negligible.
However, the usual limits can not be directly applied for the
new gauge boson couples weakly
(through its mixing with the standard $Z$ boson)
to charged fermions and the
induced flavour violation is proportional to the tiny neutrino masses
\footnote{Such an effective RH neutral coupling requires a relatively
large neutrino mixing with the new light fermions, as well as
a relatively large gauge boson mixing.
Even for a slight deviation of
$\varepsilon^{C}_{L}\ U_{\alpha i}^{*}$ in Eq. (\ref{L-int-CC-general})
from unitarity,
$\varepsilon^{C}_{L} = 0.98$, the former can be a priori as large as 0.2.
However, the latter is strongly constrained by the bound on
the $\rho$ parameter.
At any rate, the experimental limits on this $Z'$ are weaker
than the typical ones on extra gauge bosons \cite{PDG}, as
they are its couplings to standard fermions \cite{a_acl}.}.
Obviously, a detailed analysis of the present experimental
constraints from precise electroweak data \cite{PDG} is necessary
to decide on the allowed region of parameters.
Such a study is beyond the scope of this paper.

Our generic conclusion is that we should be aware of
the possibility that NP may allow to differentiate between
Dirac and Majorana neutrinos propagating in matter, and
that it is worth to search for this difference in the next
generation of very precise neutrino oscillation experiments.
A positive signal would not only indicate the Majorana character
of massive neutrinos but the existance of new interactions.
\vspace{5mm}

\noindent
\textbf{Ackowledgments}: We thank A. Bueno and J. Santiago for a careful
reading of the manuscript and useful comments.
This work has been partially supported by
Polish Ministry of Science under Grant 1 P03 B 049 26,
by MEC (FPA2006-05294), by Junta de Andaluc\'{\i}a (FQM 101 and FQM 437),
and by the European Community's Marie-Curie Research Training
Network under contract MRTN-CT-2006-035505 ``Tools and Precision
Calculations for Physics Discoveries at Colliders''.

\end{document}